\documentclass[useAMS,usenatbib]{mn2e} 
\usepackage{amsmath}
\usepackage{graphicx} 
\usepackage{aas_macros}
\usepackage[usenames]{color} 

\title[The hot $\gamma$~Doradus and Maia stars]
{The hot $\gamma$~Doradus and Maia stars}
\author[]{L. A. Balona$^1$, C. A. Engelbrecht$^2$, Y. C. Joshi$^3$, S. Joshi$^3$, 
K. Sharma$^4$, E. Semenko$^5$, 
\newauthor{G. Pandey$^6$, N. K. Chakradhari$^7$, 
David Mkrtichian$^8$, B. P. Hema$^6$, J. M. Nemec$^9$}\\
$^1$South African Astronomical Observatory, P.O. Box 9, Observatory 7935,
South Africa, lab@saao.ac.za\\
$^2$Department of Physics, University of Johannesburg, PO Box 524, Auckland
Park, Johannesburg 2006, South Africa\\
$^3$Aryabhatta Research Institute of Observational Sciences (ARIES), Manora
peak, Nainital, India\\
$^4$Department of Physics and Astrophysics, University of Delhi, Delhi - 110007,
India\\
$^5$Special Astrophysical Observatory, Russian Academy of Sciences, Nizhny
Arkhyz, 369167, Russia sea@sao.ru\\
$^6$Indian Institute of Astrophysics, Bengaluru, Karnataka 560034, India\\
$^7$School of Studies in Physics and Astrophysics, Pt Ravishankar Shukla 
University, Raipur 492 010, India\\
$^8$National Astronomical Research Institute of Thailand, 191 Huay Kaew Road, 
Muang, 50200, Chiangmai, Thailand\\
$^9$Department of Physics \& Astronomy, Camosun College, Victoria, British 
Columbia, V8P 5J2, Canada
}

\date{}

\def\LaTeX{L\kern-.36em\raise.3ex\hbox{a}\kern-.15em 
 T\kern-.1667em\lower.7ex\hbox{E}\kern-.125emX} 
 
\begin{document}

\maketitle 
 
\begin{abstract}
The hot $\gamma$~Doradus stars have multiple low frequencies characteristic
of $\gamma$~Dor or SPB variables, but are located between the red edge of
the SPB and the blue edge of the $\gamma$~Dor instability strips where all
low-frequency modes are stable in current models of these stars.  Though
$\delta$~Sct stars also have low frequencies, there is no sign of high
frequencies in hot $\gamma$~Dor stars.  We obtained spectra to refine the
locations of some of these stars in the H-R diagram and conclude that these
are, indeed, anomalous pulsating stars.  The Maia variables have multiple
high frequencies characteristic of $\beta$~Cep and $\delta$~Sct stars, but
lie between the red edge of the $\beta$~Cep and the blue edge of the
$\delta$~Sct instability strips.  We compile a list of all Maia candidates 
and obtain spectra of two of these stars.  Again, it seems likely that these 
are anomalous pulsating stars which are currently not understood.
\end{abstract} 
 

\section{Introduction}

The $\gamma$~Doradus variables are early F or late A stars which lie on or 
just above the main sequence and which vary in light with multiple frequencies
in the range 0.3--5.0\,d$^{-1}$.  The prototype of the class was discovered to
be variable by \citet{Cousins1963} and subsequently recognized as a new class 
of pulsating star by \citet{Balona1994b}. There are over 60 known $\gamma$~Dor
stars discovered by ground-based observations \citep{Handler1999, Henry2005a, 
Henry2005b, DeCat2006}.  Many hundreds more $\gamma$~Dor stars have been
discovered with the {\it Kepler} satellite \citep{Balona2011f, Bradley2015}.

The closely related $\delta$~Scuti variables are A--F stars which pulsate with 
multiple frequencies as high as 100\,d$^{-1}$, but also include the low 
frequencies found in $\gamma$~Dor stars.  Ground based observations are not
sufficiently precise to detect these low frequencies, which is why
$\delta$~Sct stars were originally thought to pulsate only in high-frequency
modes, typically with frequencies $\nu > 5$\,d$^{-1}$.  Observations from
space using the {\it Kepler} satellite show that, in fact, all $\delta$~Sct 
stars contain low frequencies \citep{Balona2014a}.  The low frequencies reach 
maximum amplitudes in the region overlapping the blue edge of the $\gamma$~Dor
instability strip.  From the ground, $\delta$~Sct stars with both high and
low frequencies, the ``$\gamma$~Dor/$\delta$~Sct hybrids'', are mostly seen
in this region of the H-R diagram.  This term arises from a misconception
since low frequencies are nearly always detected in $\delta$~Sct stars given 
sufficient precision.

Whereas $\delta$~Sct stars contain both low and high frequencies, we define a 
$\gamma$~Dor star by the presence of low frequencies and the absence of high
frequencies.  Pulsation in $\gamma$~Dor stars is driven by the  convective 
blocking mechanism \citep{Guzik2000}.  Since this requires a convective 
envelope of sufficient depth, this mechanism cannot operate in stars hotter 
than late A or early F.  The high-frequency pulsations in $\delta$~Sct stars 
are driven by the $\kappa$ (opacity) mechanism operating in the He\,II 
ionization zone \citep{Pamyatnykh2000}.  Models of non-rotating $\delta$~Sct 
stars indicate that only modes with frequencies higher than about 5\,d$^{-1}$ 
are driven.  The mechanism driving the low frequencies is currently not understood 
\citep{Balona2015d}.

The location of {\it Kepler} $\gamma$~Dor stars in the H-R diagram was 
investigated by \citet{Balona2011f}.  The blue edge of this sample of stars
is well defined and agrees with the predictions of the convective blocking
mechanism.  It should be noted, however, that is often impossible to
distinguish between pulsation and rotation for stars with only a few close
peaks, and the sample is certainly contaminated with many non-pulsating
spotted stars.  It is thought that most cool stars, either fully convective or
with a convective envelope like the Sun, will have spots on their surfaces.  
Over 500 stars of spectral types F--M are classified as rotational variables 
of this type \citep{Strassmeier2009}.  

While the vast majority of $\gamma$~Dor stars observed by {\it Kepler} are 
cooler than the granulation boundary (the division between stars with
convective and radiative envelopes), as expected, there are a few anomalous 
stars observed by {\it Kepler} which have multiple low frequencies and no 
detectable high frequencies, but seem to be significantly hotter than the 
blue edge of the $\gamma$~Dor instability strip.  These hot $\gamma$~Dor-like 
stars are discussed in \citet{Balona2014a}.  Note that \citet{Bradley2015} 
analysed a large number of faint {\it Kepler} targets and also found a few 
hot $\gamma$~Dor candidates.  

The presence of anomalous hot $\gamma$~Dor variables clearly poses a problem 
as there is no known mechanism to drive these pulsations.  The effective
temperatures, $T_{\rm eff}$, are derived from multiband Sloan photometry and 
listed in the  {\it Kepler Input Catalogue} (KIC, \citealt{Brown2011a}).  
However, the KIC photometry does not include measurements in the UV and the 
derived $T_{\rm eff}$ are not reliable for B-type stars.  It is possible that 
the $T_{\rm eff}$ for anomalous hot $\gamma$~Dor stars are in error.  They 
could, in fact, be normal $\gamma$~Dor stars or SPB stars.  The SPB variables 
are mid- to late-B stars which pulsate in multiple low frequencies driven by 
the $\kappa$~mechanism due to the opacity bump of iron-group elements. Their 
light curves closely resemble those of $\gamma$~Dor stars.

The cool edge of the SPB instability strip is at about 11500\,K and the hot 
edge of the $\delta$~Sct instability strip is at about 8500\,K.  Between
these two instability strips no pulsating star is expected to be found.  
However, there have been persistent reports of stars with high frequencies,
typical of SPB or $\delta$~Sct variables, located in this region of the H-R
diagram and termed ``Maia variables''.  \citet{Mowlavi2013} have
found what appears to be several examples of this class in photometric 
observations of the young open cluster NGC\,3766.  They discovered a large 
population (36 stars) of new variable stars between the red edge of the SPB 
instability strip and the blue edge of the $\delta$~Sct instability strip.  
Most stars have periods in the range 0.1--0.7\,d, with amplitudes between 
1--4 mmag.  About 20\,percent of stars in this region of the H-R diagram were 
found to be variable.  More recently, \citet{Balona2015c} have re-analysed the 
{\it Kepler} B stars and found many stars with well-determined $T_{\rm eff}$ 
in this region of the H-R diagram.

Apart from the low frequencies in $\delta$~Sct stars,  the hot $\gamma$~Dor 
stars and the Maia variables provide clear challenges to our current 
understanding of stellar pulsation.  The aim of this paper is to report 
and analyse new spectroscopic observations of some of these stars in the
{\it Kepler} field.  Our aim is to determine whether or not the effective 
temperatures of these stars are in error and to decide whether they may 
possibly be explained as composite stars.

\begin{figure}
\centering
\includegraphics{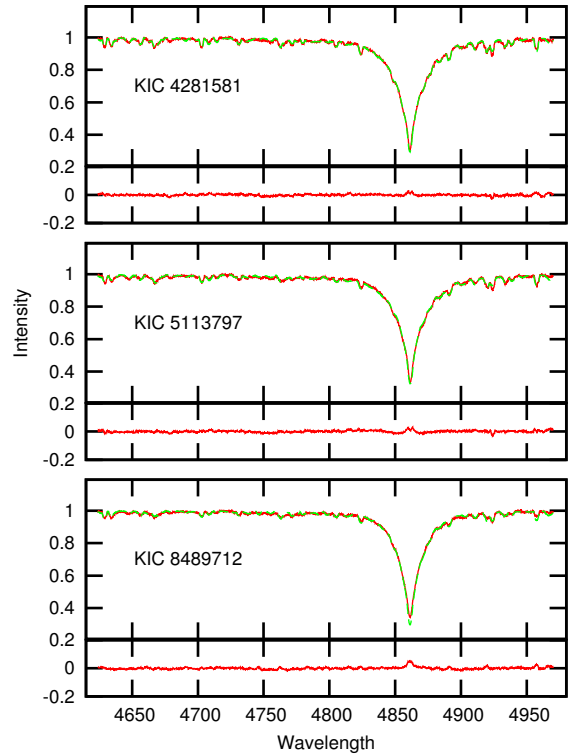}
\caption{Observed and fitted spectra in the vicinity of H$\beta$ for three 
stars observed at SAO.  The continuum intensity has been normalized to
unity.}
\label{mss}
\end{figure}

\begin{figure}
\centering
\includegraphics{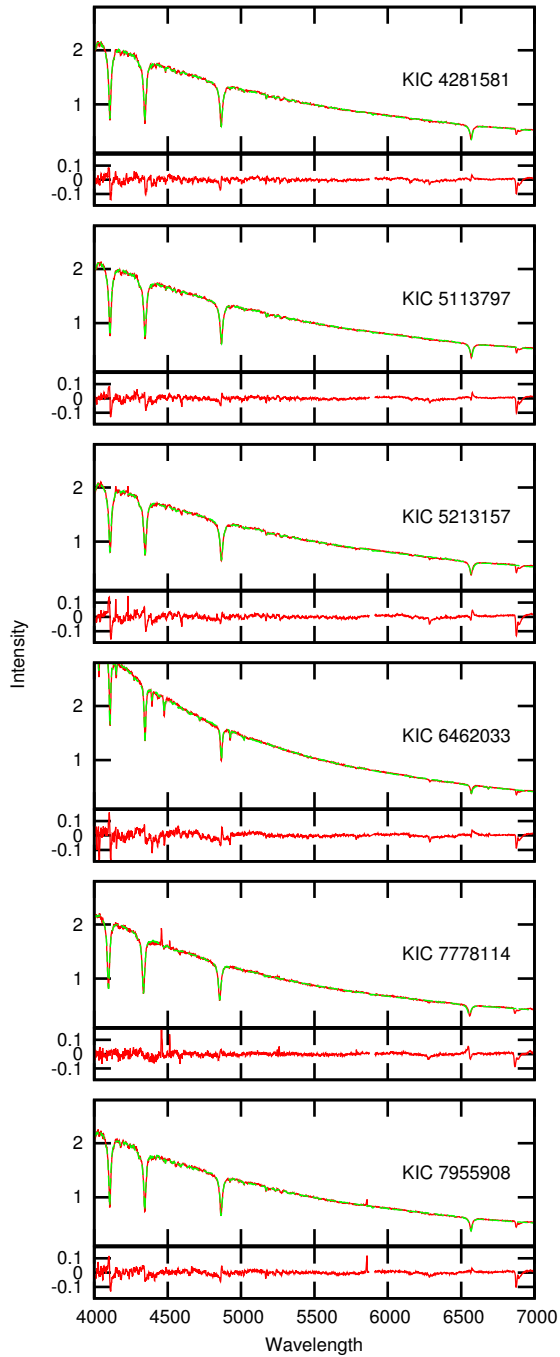}
\caption{Observed and fitted spectra (top panel) and residuals (bottom
panel) for stars observed with HFOSC.  The intensity is in arbitrary units. }
\label{hct1}
\end{figure}

\begin{figure}
\centering
\includegraphics{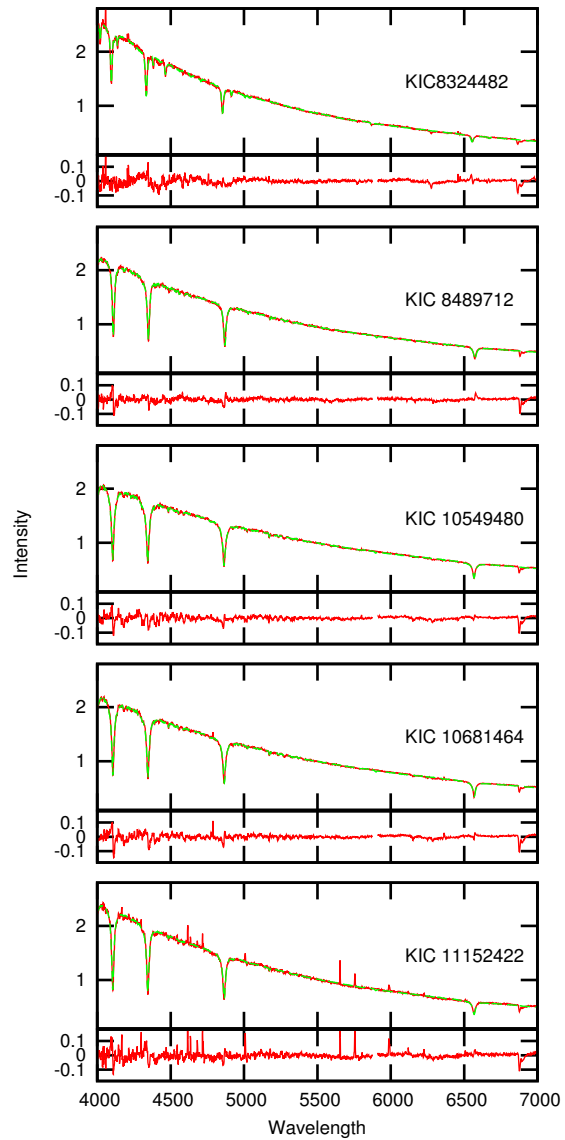}
\caption{Observed and fitted spectra (top panel) and residuals (bottom
panel) for stars observed with HFOSC.}
\label{hct2}
\end{figure}

\section{Spectroscopic observations}

Spectra for three stars, KIC\,4281581, KIC\,5113797 and KIC\,8489712, 
were obtained in 2014 March at the Special Astrophysical Observatory (SAO) of 
the Russian Academy of Sciences with the MSS spectrograph attached to the 6-m 
telescope (BTA).  The detector is a 2K$\times$4K CCD.  The spectra cover the 
wavelength range 4430--4985\,{\AA} with $R = 14000$.  

The analysis was carried out in a semi-automatic manner with the latest 
version of SME \citep{Valenti1996}.  It is sometimes difficult to find a
suitable fit to the core of the hydrogen line profiles, which is quite
sensitive to the effective temperature and the projected rotational
velocity.  Observed and fitted spectra of these stars are shown in 
Fig.\,\ref{mss}.  

Low resolution spectra were obtained on 2015 September 23--24 with the 
Himalaya Faint Object Spectrograph and Camera (HFOSC) mounted on the
2.0-m Himalayan Chandra Telescope (HCT) operated by the Indian Institute of 
Astrophysics (IIA), Bangalore.  The detector is a 2K$\times$2K SITe CCD.
Most of the spectra cover the wavelength range 3800--7000\,{\AA} with a 
spectral resolution $R = 1300$.  A few spectra were obtained in the wavelength
region 5800--8350\,{\AA} with $R = 2190$.

The spectra were analysed using {\tt ULySS} (University of Lyon spectroscopic
software, \citealt{Koleva2009}) which can also be used for the determination 
of the stellar atmospheric parameters.  A series of modelled spectra is 
generated based on an empirical library and compared with the observed 
spectrum. The sum of the squares of the differences between the observed and 
modelled spectrum is minimized to obtain the best fitting parameters.
The selection of the empirical library depends upon the coverage of the 
stellar parameters and the spectral resolution.  For our analysis, the
{\tt MILES} interpolator \citep{Prugniel2011}, based on the {\tt MILES}
 library \citep{Sanchez2006} was used.  The resolution of the {\tt MILES}
spectra, $R \approx 2000$, closely matches with that of the spectra
obtained with HFOSC at HCT.  The observed and fitted spectra are shown in 
Figs.\,\ref{hct1} and \ref{hct2}.

The derived stellar parameters are listed in Table\,\ref{gdor} for those stars
that we classified as anomalous hot $\gamma$~Dor stars and in 
Table\,\ref{nongdor} for stars which could be assigned normal variability 
types, either SPB or $\delta$~Sct.

\begin{table*}
\caption{List of hot $\gamma$~Doradus stars. The spectral type is taken from 
the literature.  The {\it Kepler} magnitude, Kp, the corrected effective 
temperature,  $T_{\rm eff}^1$, and surface gravity, $\log g^1$, are from the 
KIC.  The  values of  $T_{\rm eff}$, $\log g$, [Fe/H] and projected rotational
velocity, $v \sin i$, are derived from spectra.  The source of the 
spectroscopic stellar parameters is given in the Ref column as follows: 
1 - MSS data; 2 - HFOSC data; 3 - \citet{Niemczura2015};  
4 - \citet{Tkachenko2013b}; 5 - \citet{Ulusoy2014};  6 - \citet{Antonello2006};
7 - \citet{McDonald2012};  no reference - KIC photometric values.  The 
luminosity, $\log L/L_\odot$, is derived using \citet{Torres2010a}.  Possible 
rotation periods, derived from the light curve, are given.  The references for 
these periods are: a - \citet{Nielsen2013};  b - \citet{Reinhold2013}; 
c -\citet{Balona2013c}.}  

\label{gdor}
\begin{tabular}{rrrrrrrrrrrl}
\hline
\multicolumn{1}{c}{KIC} & 
\multicolumn{1}{c}{Sp.} &
\multicolumn{1}{c}{Kp} &
\multicolumn{1}{c}{$T_{\rm eff}^1$} &
\multicolumn{1}{c}{$\log g^1$} &
\multicolumn{1}{c}{$T_{\rm eff}$} &
\multicolumn{1}{c}{$\log g$} &
\multicolumn{1}{c}{[Fe/H]} &
\multicolumn{1}{c}{$v \sin i$} &
\multicolumn{1}{c}{Ref} &
\multicolumn{1}{c}{$\log \frac{L}{L_\odot}$} & 
\multicolumn{1}{c}{Period} \\
\multicolumn{1}{c}{} & 
\multicolumn{1}{c}{type} & 
\multicolumn{1}{c}{mag} &
\multicolumn{1}{c}{K} &
\multicolumn{1}{c}{dex} &
\multicolumn{1}{c}{K} &
\multicolumn{1}{c}{dex} &
\multicolumn{1}{c}{dex} & 
\multicolumn{1}{c}{km\,s$^{-1}$} & 
\multicolumn{1}{c}{} & 
\multicolumn{1}{c}{} & 
\multicolumn{1}{c}{d} \\
\hline
  4934767 &          & 11.24  & 9870 &  4.25  &       &      &       &     &   &  1.44 &             \\
  5113797 &          &  9.15  & 8280 &  3.83  &  7710 & 3.5  & -0.3  & 137 & 1 &  1.83 & 2.418$^a$   \\
          &          &        &      &        &  8700 & 4.5  & -0.4  &     & 2 &  0.90 &             \\
          & A3IV-V   &        &      &        &  8100 & 4.0  &       & 112 & 3 &       &             \\
  5213157 &          & 11.94  & 8460 &  3.81  &  8700 & 4.5  & -0.3  &     & 2 &  0.90 & 0.5452$^b$  \\
  5429163 & A5V      &  9.72  & 8220 &  4.02  &  8100 & 4.0  & -0.1  & 163 & 3 &  1.31 &             \\
  7694191 &          & 10.78  & 8000 &  3.54  &       &      &       &     &   &  1.83 & 0.3762$^c$  \\
  7767565 & Am       &  9.32  &      &        &  7800 & 3.8  &  0.4  &  65 & 3 &  1.51 &             \\
  7778114 &          & 11.98  & 8890 &  4.11  &  9500 & 4.1  & -0.5  &     & 2 &  1.50 &             \\ 
  8489712 & A2IVs    &  8.62  & 8490 &  3.52  &  8150 & 3.0  & -0.3  & 120 & 1 &  2.60 &             \\
          &          &        &      &        &  8650 & 3.4  & -0.7  &     & 2 &  2.20 &             \\
          &          &        &      &        &  8800 & 3.5  &  0.1  & 126 & 3 &       &             \\ 
          &          &        &      &        &  8270 & 2.9  & -0.6  & 119 & 4 &       &             \\
  8523871 &          & 12.36  & 8390 &  4.45  &       &      &       &     &   &  0.84 &             \\   
 10096499 & A3V      &  6.92  & 7920 &  4.13  &  7960 & 3.3  &  0.0  &  90 & 4 &  2.21 &             \\
 10549480 & A2       &  9.75  & 8580 &  3.95  &  8900 & 4.6  & -0.4  &     & 2 &  0.80 & 0.9579$^c$  \\
 10681464 &          & 11.48  & 8970 &  4.02  &  8900 & 4.5  & -0.3  &     & 2 &  1.00 & 0.8599$^b$  \\
HD 148542 & A3IV     &  6.03  & 8470 &        &       &      &       &  92 & 6,7&  2.04 &            \\
\hline                                                                                                
\end{tabular}
\end{table*}                                                                                          

\begin{table*}
\caption{Other stars observed spectroscopically. The columns are the same as 
in Table\,\ref{gdor} except we have added a possible variability type in the
last column.}
\label{nongdor}
\begin{tabular}{rrrrrrrrrrrll}
\hline
\multicolumn{1}{c}{KIC} & 
\multicolumn{1}{c}{Sp.} &
\multicolumn{1}{c}{Kp} &
\multicolumn{1}{c}{$T_{\rm eff}^1$} &
\multicolumn{1}{c}{$\log g^1$} &
\multicolumn{1}{c}{$T_{\rm eff}$} &
\multicolumn{1}{c}{$\log g$} &
\multicolumn{1}{c}{[Fe/H]} &
\multicolumn{1}{c}{$v \sin i$} &
\multicolumn{1}{c}{Ref} &
\multicolumn{1}{c}{$\log \frac{L}{L_\odot}$} & 
\multicolumn{1}{c}{Period} &
\multicolumn{1}{c}{Type} \\
\multicolumn{1}{c}{} & 
\multicolumn{1}{c}{type} & 
\multicolumn{1}{c}{mag} &
\multicolumn{1}{c}{K} &
\multicolumn{1}{c}{dex} &
\multicolumn{1}{c}{K} &
\multicolumn{1}{c}{dex} &
\multicolumn{1}{c}{dex} & 
\multicolumn{1}{c}{km\,s$^{-1}$} & 
\multicolumn{1}{c}{} & 
\multicolumn{1}{c}{} & 
\multicolumn{1}{c}{d} &
\multicolumn{1}{c}{} \\
\hline
  3756846 &          & 15.75  &11000 &  3.94  &       &      &       &     &   &  2.08 &            & SPB   \\
  4281581 &          &  9.40  & 8290 &  3.84  &  8090 & 3.5  & -0.1  & 113 & 1 &  1.93 &            & $\delta$~Sct? \\
          &          &        &      &        &  8900 & 4.5  & -0.3  &     & 2 &  1.00 &            &       \\
          & A3IV-Vs  &        &      &        &  8200 & 3.9  &       & 105 & 3 &       &            &       \\
  6462033 &          & 10.72  & 8530 &  4.32  & 18400 & 4.2  & -0.3  &     & 2 &  3.00 & 0.6994$^b$ & SPB   \\
          &          &        &      &        &  7150 & 4.3  &       &  90 & 5 &       &            &       \\            
  7955898 &          & 12.32  & 8680 &  3.65  &  8300 & 3.7  & -0.5  &     & 2 &  1.70 & 0.8132$^b$ & $\delta$~Sct? \\
  8324482 & A0       & 11.61  & 8300 &  3.82  & 18900 & 4.2  & -0.5  &     & 2 &  3.00 &            & SPB   \\
 11152422 &          & 15.00  & 9290 &  4.02  &  9000 & 4.0  & -1.4  &     & 2 &  1.40 &            & $\delta$~Sct? \\
\hline                                                                                                
\end{tabular}
\end{table*}

\section{Stellar parameters}

It should be noted that the KIC effective temperatures for A stars are 
remarkably consistent with those derived from high-dispersion spectroscopy.
From 107 stars with both KIC and spectroscopic parameters, \citet{Balona2015d} 
found  “ that spectroscopic estimates of $T_{\rm eff}$ of A stars are 
systematically 144\,K higher than those in the KIC (the corrected KIC values 
are shown in Tables\,\ref{gdor} and \ref{nongdor}).  Moreover, the standard 
deviation for $T_{\rm eff}$ is 150\,K from spectroscopy and 200\,K for the KIC
value.  The KIC effective temperature is thus quite sufficient to distinguish 
between normal $\gamma$~Dor stars and hot $\gamma$~Dor variables.  However, 
there is a problem with the KIC value of $T_{\rm eff}$ for the B stars. B 
stars can easily be assigned temperatures within the A star region.  What is 
required is simply to distinguish between an A star and a B star, and this is 
easily done with low-dispersion spectra by looking for the presence of He\,I 
lines.

Stellar parameters determined from the low-dispersion HFOSC spectra have 
typical errors as follows: $T_{\rm eff} = 300$\,K,  $\log g = 0.17$\,dex,
[Fe/H] = 0.13.  Parameters from the MSS observations have the following
typical errors: $T_{\rm eff} = 250$\,K,  $\log g = 0.15$\,dex, [Fe/H] = 0.1.
These are certainly sufficient to locate a star in the H-R diagram with 
enough precision for our purposes.  In addition to our spectroscopic
results, Tables\,\ref{gdor} and \ref{nongdor} also list spectroscopic 
parameters derived from the literature.

\section{The hot $\gamma$~Dor stars}

Most of the hot $\gamma$~Dor star candidates listed in Table\,\ref{gdor} are
from \citet{Balona2014a}.  Also shown are additional candidates selected from 
{\it Kepler} $\gamma$~Dor stars that are hotter than the granulation 
boundary.  \citet{Antonello2006} discovered line profile variations in 
HD\,148542 (HR\,6139) that can be explained by prograde g-modes similar to 
those observed in SPB stars, though it has a spectral type of A2V.  This star 
can also be considered a candidate hot $\gamma$~Dor variable.

HD\,208727 was reported as an anomalous pulsator, but \citet{Kallinger2002} 
found it to be a rotational variable with a period of 0.317\,d.  Other 
candidates such as HD\,29573 and $\gamma$~CrB \citep{Percy2000} are most 
likely rotational variables as well, but further observations are required.

\begin{figure}
\centering
\includegraphics{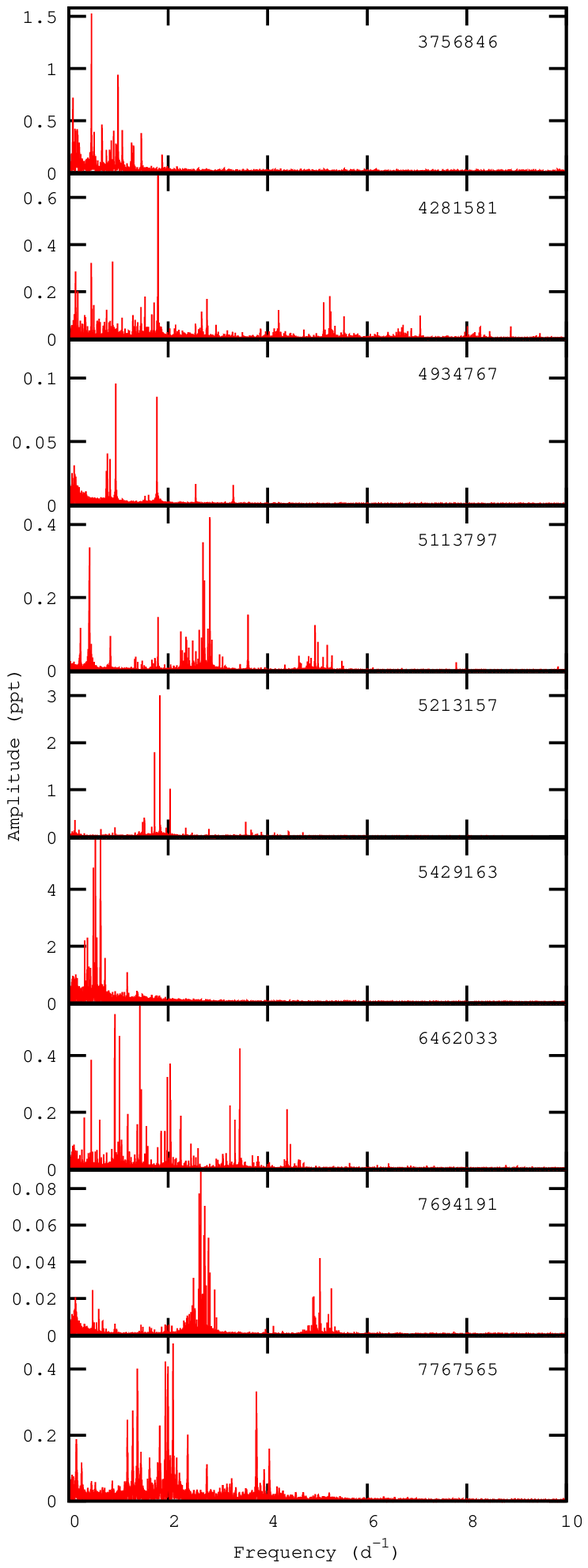}
\caption{Periodograms of the hot $\gamma$~Dor stars from KIC photometry.
The stars are the anomalous hot $\gamma$~Dor stars listed in 
\citet{Balona2014a}.}
\label{period1}
\end{figure}

\begin{figure}
\centering
\includegraphics{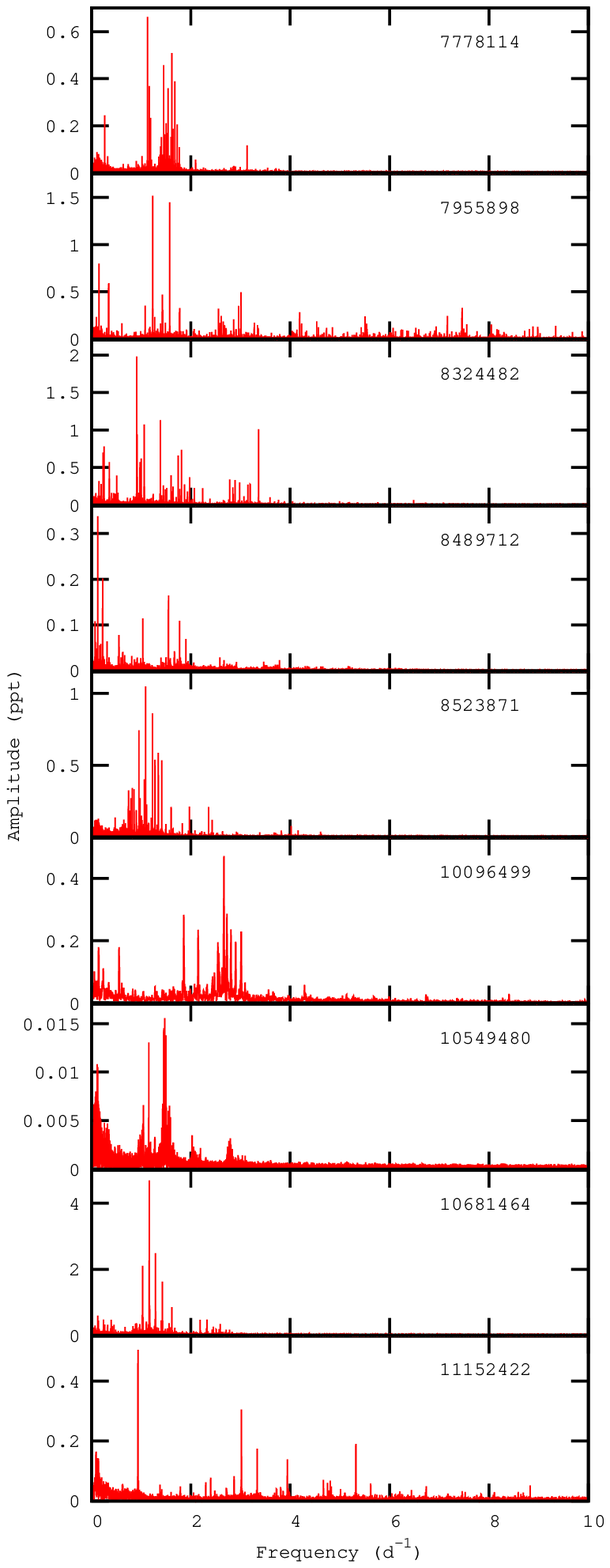}
\caption{Periodograms of the hot $\gamma$~Dor stars from KIC photometry.
The stars are the anomalous hot $\gamma$~Dor stars listed in 
\citet{Balona2014a}.}
\label{period2}
\end{figure}

The periodograms of the long-cadence (30-min exposures) of these stars and
other stars are shown in Figs.\,\ref{period1} and \ref{period2}.  Note that in 
long-cadence data a frequency, $\nu$, above 24\,d$^{-1}$ appears as a peak at 
approximately $(48 - \nu)$\,d$^{-1}$.  If the data were taken at strictly exact 
time intervals, it would not be possible to discern any frequencies higher 
than 24\,d$^{-1}$. Fortunately this is not the case and it is 
indeed possible to identify frequencies higher than this value if the 
amplitudes are sufficiently high \citep{Murphy2013a}.  Examination of the 
long-cadence periodograms show that no such peaks are visible.  Moreover, 
short-cadence data are available for KIC\,5113797, 8489712 and 10549480 and 
it is clear that high frequencies are absent in these three stars.

In Table\,\ref{nongdor} we show stars which are probably SPB or $\delta$~Sct
variables. KIC\,4281581, KIC\,7955898 and KIC\,11152422 have some very 
low-amplitude peaks in the high-frequency range (Figs.\,\ref{period1}, 
\ref{period2}) and are possibly $\delta$~Sct variables.  KIC\,3756846, 
6462033, 8324482 are hot enough to be classified as SPB stars.

For KIC\,8489712 there is quite a large discrepancy between the [Fe/H] 
value from \citet{Tkachenko2013b} and from \citet{Niemczura2015}.  The 
difference in line strengths between the two abundance values is large,
so either there is a mistake in one of these measurements or the 
star is a spectrum variable.  Our measurements support the lower metallicity.

KIC\,5113797, which is close to the open cluster NGC 6819, appears to show one
flare \citep{Balona2015a} and is also an X-ray source \citep{Gosnell2012}. 

The largest discrepancy in Table\,\ref{nongdor} occurs for KIC\,6462033 where
we find $T_{\rm eff} = 18400$\,K, but \citet{Ulusoy2014} find $T_{\rm eff} =
7150$\,K.  Our spectrum shows strong He\,I\,4437 and He\,I\,4471 which can only
occur in a B star, so the difference is very puzzling.  The field is
slightly crowded, but KIC\,6462033 is by far the brightest star in the
field.  The available photometry gives $B-V = -0.05$ which would also
indicate a B star rather than a mid-A star.

\begin{figure}
\centering
\includegraphics{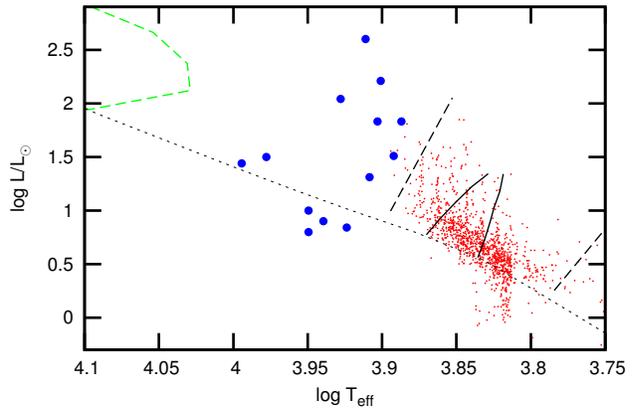}
\caption{The location of the hot $\gamma$~Dor variables in Table\,\ref{gdor}
in the theoretical H-R diagram is shown by large filled circles.  The small 
(red) dots are normal $\gamma$~Dor stars in the {\it Kepler} field.   The 
dashed (green) region in the top left corner is the cool end of the SPB 
instability strip.  The red and blue edges of the $\gamma$~Dor instability 
strip calculated by \citet{Dupret2004a} are shown as solid lines, while
those calculated by \citet{Xiong2016} are shown by the dashed lines. The 
dotted line is the zero-age main sequence for solar composition (from 
\citealt{Bertelli2008}).}
\label{hrdiag}
\end{figure}

The results in Table\,\ref{gdor} mostly confirm that the stars are
indeed significantly hotter than the blue edge of the known $\gamma$~Dor 
stars.  Using values of  $T_{\rm eff}$, $\log g$ and [Fe/H], the luminosity 
may be estimated using the relationship in \citet{Torres2010a}. For those 
stars without spectroscopic estimates, we use the values listed in the KIC.
The location of these stars in the theoretical H-R diagram is shown in 
Fig.\,\ref{hrdiag}.  The red and blue edges of the $\gamma$~Dor instability
strips shown in this figure were both calculated using time-dependent theories 
of convection.  In the calculations by \citet{Dupret2004a}, the location of 
the instability strip depends on the choice of mixing length parameter, 
$\alpha$.  Here we show the edges calculated with $\alpha = 2.0$ for modes 
with spherical harmonic $l = 1$.  The strip moves towards cooler temperatures 
for smaller values of $\alpha$. The red and blue edges calculated by 
\citet{Xiong2016} are for $l \le 4$ and use a different non-local, 
time-dependent convection theory.  The differences in location of the red
and blue edges is quite remarkable and illustrates our poor understanding of 
convection and how it interacts with pulsation.  As can be seen, the hot 
$\gamma$~Dor stars are located outside the blue edge of the  $\gamma$~Dor 
instability strip and the red edge of the SPB instability strip.

\section{Maia variables}

Reports of pulsating stars between the blue edge of the $\delta$~Sct
instability strip and the red edge of the SPB instability strip have been 
reported from time to time.  \citet{Struve1955} suggested the possibility of 
a new class of variables on the basis of radial velocity observations of the 
late B star Maia, a member of the Pleiades.  Later, \citet{Struve1957} 
disclaimed such variability in Maia.  Since that time, several attempts have 
been made to detect periodic variability in stars lying in this region of the 
H-R diagram, without success \citep{McNamara1985, Scholz1998, Lehmann1995}.
The situation is confusing because in these searches no distinction was made 
between high and low frequencies.  Some variable stars lying in this region 
are probably best explained as rotational variables, as mentioned above.  

Pulsation at high frequencies (i.e. higher than about 5\,d$^{-1}$) are
expected in the $\delta$~Sct stars but also among the $\beta$~Cep variables. 
These are typically O9--B3 stars where the pulsations are driven by the
ionization of metals of the iron group elements \citep{Dziembowski1993a}.
We define Maia variables as stars lying between the red edge of the 
$\beta$~Cep instability strip and the blue edge of the $\delta$~Sct instability
strip which show multiple high frequencies.  Models indicate that such stars 
should not exist in this region. 

One such star, $\alpha$~Dra, is classified as A0\,III with a period of about 
53\,min and very small amplitude.  This may be considered as a good example of
a Maia variable \citep{Kallinger2004}.  \citet{Degroote2009b} and 
\citet{Balona2011b} found examples of B-type stars with high frequencies and 
low amplitudes which are significantly cooler than the red edge of the 
$\beta$~Cep instability strip.  These may also be considered as Maia
candidates.  

A list of Maia candidates compiled from the literature is given in 
Table\,\ref{maia}.  The atmospheric parameters of the presumed Maia
variables observed by {\it CoRoT}, (CoRoT 102729531, 102771057, 102790063, 
102861067, \citealt{Degroote2009b}) were obtained from Str\"{o}mgren photometry
without the $\beta$ index.  Subsequently, spectroscopic observations have 
shown that these are mid-A stars \citep{Sebastian2012}. Therefore they are
probably just normal $\delta$~Sct variables. Table\,\ref{maia} lists the
five remaining {\it CoRoT} Maia candidates without spectral classifications.
Better estimates of the stellar parameters are required to decide whether
they are Maia variables or normal $\delta$~Sct stars.

\begin{table}
\caption{List of Maia star candidates.  The spectral type and magnitude are
shown together with estimates of the effective temperature, $T_{\rm eff}$
luminosity $\log L/L_\odot$, and projected rotational velocity.  The last
column shows the reference as follows: 1 - \citet{Prugniel2011};
2 - \citet{Balona2015c}; 3 - \citet{Degroote2009b}; 4 - \citet{Aerts2005};
5 - \citet{Mowlavi2013}; 6 - \citet{Lata2014}.}
\resizebox{0.85\textwidth}{!}{\begin{minipage}{\textwidth}
\label{maia}
\begin{tabular}{llrrrrl}
\hline
\multicolumn{1}{c}{Name} & 
\multicolumn{1}{c}{Sp.} &
\multicolumn{1}{c}{V} &
\multicolumn{1}{c}{$T_{\rm eff}$} &
\multicolumn{1}{c}{$\log \frac{L}{L_\odot}$} &
\multicolumn{1}{c}{$v \sin i$} &
\multicolumn{1}{l}{Ref} \\
\multicolumn{1}{c}{} & 
\multicolumn{1}{c}{type} & 
\multicolumn{1}{c}{mag} &
\multicolumn{1}{c}{K} &
\multicolumn{1}{c}{   } &
\multicolumn{1}{c}{km\,s$^{-1}$} & 
\multicolumn{1}{c}{} \\
\hline
$\alpha$~Dra    & A0III     &  3.68 & 10100 &  2.68 &   26 & 1   \\
HD 121190       & B9V       &       & 12200 &  1.96 &  118 & 4,5 \\
HD 189637       &           & 10.86 & 19000 &  2.96 &      & 2   \\
HD 234893       & B5V       &  9.33 & 14700 &  2.87 &  130 & 2   \\
HD 234999       & B9        &  9.09 & 11150 &  2.08 &  103 & 2   \\                                                             
HD 251584       & B9        & 11.12 &       &       &      & 2   \\
HD 253107       & B1V       & 10.37 & 17600 &       &      & 2   \\
CoRoT 102862454 &           &       &       &       &      & 3   \\
CoRoT 102790331 &           &       &       &       &      & 3   \\
CoRoT 102848985 &           &       &       &       &      & 3   \\
CoRoT 102922479 &           &       &       &       &      & 3   \\
CoRoT 102850576 &           &       &       &       &      & 3   \\
CoRoT 102850502 &           &       &       &       &      & 3   \\
KIC 2987640     &           & 12.64 & 11000 &  1.87 &      & 2   \\
KIC 3343239     &           & 14.42 & 10691 &  1.97 &      & 2   \\
KIC 3459297     &           & 12.55 & 10600 &  1.14 &      & 2   \\
KIC 3756031     & B5IV-V    & 10.00 & 16000 &  3.21 &   31 & 2   \\
KIC 4939281     &           & 12.08 & 18900 &  3.08 &      & 2   \\
KIC 8714886     &           & 10.95 & 19000 &  2.96 &      & 2   \\
KIC 10285114    &           & 11.23 & 18200 &  2.74 &      & 2   \\
KIC 11454304    &           & 12.95 & 17500 &  3.22 &      & 2   \\
KIC 11973705    & B8.5V/IV  &  9.12 & 11150 &  2.11 &  103 & 2   \\
KIC 12258330    & B5V       &  9.52 & 14700 &  2.88 &  130 & 2   \\
EPIC 202061002  & B9        & 12.40 &  9790 &  1.09 &      & 2   \\
EPIC 202061129  & B9        & 14.01 &  9600 &       &      & 2   \\
EPIC 202061131  & B5        & 14.23 & 14200 &       &      & 2   \\
EPIC 202062129  & B..       & 11.62 & 18000 &       &      & 2   \\
\hline                                                                                                
\end{tabular}
\end{minipage} }
\end{table}                                                                                          

\setcounter{table}{2}
\begin{table}
\caption{Continued.}
\resizebox{0.85\textwidth}{!}{\begin{minipage}{\textwidth}
\begin{tabular}{llrrrrl}
\hline
\multicolumn{1}{c}{Name} & 
\multicolumn{1}{c}{Sp.} &
\multicolumn{1}{c}{V} &
\multicolumn{1}{c}{$T_{\rm eff}$} &
\multicolumn{1}{c}{$\log \frac{L}{L_\odot}$} &
\multicolumn{1}{c}{$v \sin i$} &
\multicolumn{1}{l}{Ref} \\
\multicolumn{1}{c}{} & 
\multicolumn{1}{c}{type} & 
\multicolumn{1}{c}{mag} &
\multicolumn{1}{c}{K} &
\multicolumn{1}{c}{   } &
\multicolumn{1}{c}{km\,s$^{-1}$} & 
\multicolumn{1}{c}{} \\
\hline
NGC\,3766-50    &           & 11.08 & 16600 &  3.04 &      & 5   \\
NGC\,3766-58    &           & 11.22 & 17200 &  2.96 &      & 5   \\
NGC\,3766-59    &           & 11.25 & 16500 &  2.95 &      & 5   \\
NGC\,3766-60    &           & 11.28 & 16500 &  2.93 &      & 5   \\
NGC\,3766-62    &           & 11.39 & 16500 &  2.87 &      & 5   \\
NGC\,3766-68    &           & 11.49 & 16300 &  2.81 &      & 5   \\
NGC\,3766-69    &           & 11.50 & 15600 &  2.80 &      & 5   \\
NGC\,3766-71    &           & 11.56 & 16600 &  2.77 &      & 5   \\
NGC\,3766-78    &           & 11.67 & 16500 &  2.70 &      & 5   \\
NGC\,3766-79    &           & 11.72 & 16000 &  2.67 &      & 5   \\
NGC\,3766-83    &           & 11.77 & 15800 &  2.64 &      & 5   \\
NGC\,3766-94    &           & 11.92 & 14600 &  2.55 &      & 5   \\
NGC\,3766-105   &           & 12.11 & 15500 &  2.43 &      & 5   \\
NGC\,3766-106   &           & 12.14 & 15800 &  2.42 &      & 5   \\
NGC\,3766-112   &           & 12.20 & 15600 &  2.38 &      & 5   \\
NGC\,3766-135   &           & 12.51 & 14500 &  2.19 &      & 5   \\
NGC\,3766-136   &           & 12.53 & 13300 &  2.18 &      & 5   \\
NGC\,3766-142   &           & 12.65 & 14700 &  2.10 &      & 5   \\
NGC\,3766-144   &           & 12.61 & 12200 &  2.10 &      & 5   \\
NGC\,3766-145   &           & 12.68 & 14200 &  2.09 &      & 5   \\
NGC\,3766-147   &           & 12.74 & 12900 &  2.04 &      & 5   \\
NGC\,3766-149   &           & 12.74 & 13700 &  2.05 &      & 5   \\
NGC\,3766-161   &           & 12.86 & 13800 &  1.97 &      & 5   \\
NGC\,3766-167   &           & 12.90 & 12900 &  1.95 &      & 5   \\
NGC\,3766-170   &           & 12.92 & 14000 &  1.93 &      & 5   \\
NGC\,3766-175   &           & 12.96 & 14000 &  1.91 &      & 5   \\
NGC\,3766-194   &           & 13.10 & 13200 &  1.82 &      & 5   \\
NGC\,3766-236   &           & 13.39 & 12300 &  1.65 &      & 5   \\
NGC\,3766-259   &           & 13.56 & 12000 &  1.54 &      & 5   \\
NGC\,3766-278   &           & 13.72 & 10300 &  1.45 &      & 5   \\
NGC\,1893-V60   &           & 14.53 &  9730 &  1.61 &      & 6   \\
NGC\,1893-V70   &           & 14.51 &  9700 &  1.61 &      & 6   \\
NGC\,1893-V71   &           & 14.44 & 12400 &  1.86 &      & 6   \\
NGC\,1893-V72   &           & 14.21 &  9900 &  1.75 &      & 6   \\
NGC\,1893-V74   &           & 14.22 & 10700 &  1.82 &      & 6   \\
NGC\,1893-V79   &           & 14.71 & 11400 &  1.67 &      & 6   \\
NGC\,1893-V81   &           & 14.17 & 12300 &  1.95 &      & 6   \\
NGC\,1893-V84   &           & 14.66 & 11500 &  1.75 &      & 6   \\
NGC\,1893-V89   &           & 14.24 & 11000 &  1.86 &      & 6   \\
NGC\,1893-V90   &           & 14.65 & 11200 &  1.73 &      & 6   \\
NGC\,1893-V91   &           & 14.56 & 12000 &  1.79 &      & 6   \\
NGC\,1893-V98   &           & 14.72 & 11000 &  1.64 &      & 6   \\
NGC\,1893-V102  &           & 14.38 &  9750 &  1.65 &      & 6   \\
NGC\,1893-V109  &           & 14.61 & 11700 &  1.75 &      & 6   \\
NGC\,1893-V123  &           & 14.45 &  9620 &  1.64 &      & 6   \\
NGC\,1893-V129  &           & 14.39 &  9620 &  1.63 &      & 6   \\
NGC\,1893-V145  &           & 13.97 & 11500 &  1.99 &      & 6   \\
\hline                                                                                                
\end{tabular}
\end{minipage} }
\end{table}                                                                                          

\begin{figure}
\centering
\includegraphics{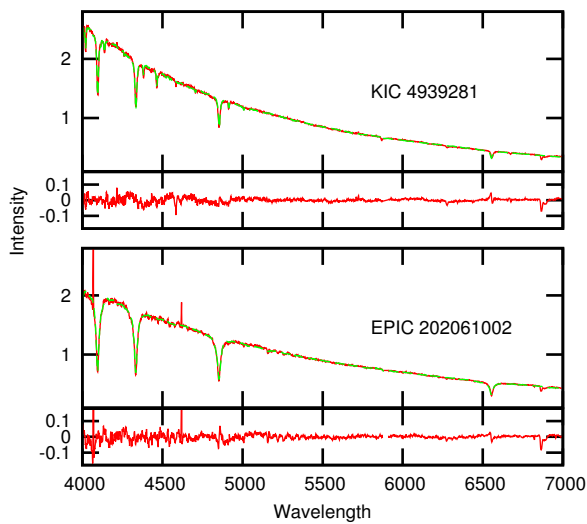}
\caption{Observed and fitted spectra (top panel) and residuals (bottom
panel) for two Maia candidates observed with HFOSC.}
\label{maiaspec}
\end{figure}

Estimation of atmospheric parameters in stars with spectral types between
late B and early A is difficult.  In this region the strength of the Balmer 
lines reaches a maximum.  Since the effective temperature is estimated by 
fitting the H$\alpha$ or H$\beta$ line profiles, the value of $T_{\rm eff}$ is 
uncertain.  For this reason, the discovery of possible Maia candidates in open
clusters, such as NGC\,3766 \citep{Mowlavi2013} and NGC\,1893 \citep{Lata2014},
is very important because the position of the star along the main sequence is 
a much better indicator of $T_{\rm eff}$. 

We observed two Maia candidates, EPIC\,202061002 and KIC\,4939281
\citep{Balona2015c} with the HFOSC.  These spectra are shown in
Fig.\,\ref{maiaspec}.  For EPIC\,202061002  the best fit gives $T_{\rm eff} =
 9790 \pm 340$, $\log g = 4.46 \pm 0.17$, [Fe/H] =$-1.7 \pm  0.1$.  For
KIC\,4939281, $T_{\rm eff} = 18900 \pm 660$, $\log g = 4.16 \pm 0.17$,
[Fe/H] =$-0.4 \pm 0.1$.  KIC\,4939281 is just outside the $\beta$~Cep
instability strip and is perhaps better classified as a $\beta$~Cep
variable.

\citet{Mowlavi2013} divides the stars in NGC\,3766 into five variability 
groups.  The 13 stars in Group 1 are mostly monoperiodic.  Although they 
classified these stars as SPB variables, they may just as well be rotational 
variables.  The 36 stars in Group 2 have periods in the range $0.1 < P < 1.1$\,d.
Of these, 23 stars are monoperiodic, 11 biperiodic and 2 have three 
frequencies.  These stars fall between the SPB and $\delta$~Sct stars and are 
thus good candidates for Maia variables.  However, the monoperiodic stars with
periods longer than about 0.5\,d can be considered as possible rotational 
variables.  In Table\,\ref{maia} those stars which have periods too short to 
be due to rotation are listed.  The other Groups consist of $\delta$~Sct and 
$\gamma$~Dor stars.

We assume a distance modulus of $V_0 - M_V = 11.6 \pm 0.2$\,mag and colour 
excess E(B-V) = 0.22\,mag for NGC\,3766 \citep{McSwain2008}.  Given the apparent
magnitude and assuming uniform reddening, we can determine the absolute 
magnitude, $M_V$.  The observed main sequence is well-defined and represents 
an isochrone of about 2.5\,Myr \citep{McSwain2008}.  We can fit an isochrone 
of this age and solar abundance using the models in \citet{Bressan2012}.  
From the isochrone we can derive $T_{\rm eff}$ as a function of colour index
and $\log L/L_\odot$ from $M_V$.  We used the bolometric correction as a 
function of $T_{\rm eff}$ by \citet{Torres2010b}.  The resulting values of 
$T_{\rm eff}$ and $\log L/L_\odot$ are shown in Table\,\ref{maia}.

Among the 104 variables in NGC\,1893 found by \citet{Lata2014}, 17 stars 
could be classified as Maia variables on the basis of their locations in the 
H-R diagram and their short periods. These stars are listed in 
Table\,\ref{maia} with effective temperatures and luminosities taken from 
their paper.  Except for a few stars with periods close to 0.5\,d, the 
periods are too short to be due to rotation. 

It should be noted that both \citet{Mowlavi2013} and \citet{Lata2014}
provide no details regarding the significance of the detected periods or the 
alias structure.  Results are presented in the form of phased light curves 
which, unfortunately, hides these details.  It would have been far better to
present the periodograms, which would enable a much more precise visual
picture of the frequency structure and noise level in these data.  

\begin{figure}
\centering
\includegraphics{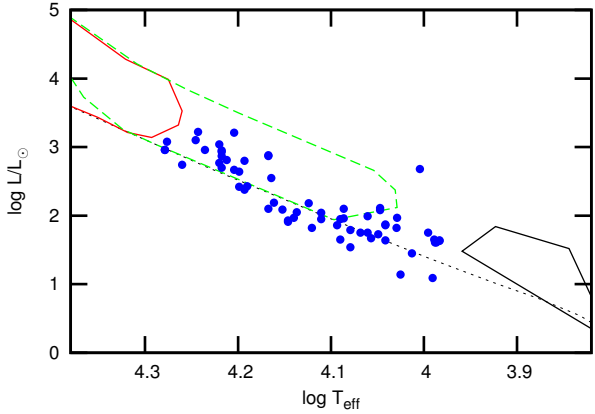}
\caption{The location of Maia candidates (Table\,\ref{maia}) in the
theoretical H-R diagram. The three regions are the $\beta$~Cep instability 
strip in the upper left corner (solid, red), the SPB instability strip 
(dashed, green) and the region populated by {\it Kepler} $\delta$~Sct 
stars in the lower right corner (solid, black). The dotted line 
is the zero-age main sequence.}
\label{hrmaia}
\end{figure}

In Fig.\,\ref{hrmaia} we show the location of the Maia candidates in the H-R
diagram with respect to the $\beta$~Cep, SPB and $\delta$~Sct instability
strips.  While it is possible that hotter stars may be re-classified as
$\beta$~Cep variables and the cooler stars as $\delta$~Sct variables, there
still remains a considerable number of stars for which such a
re-classification is unlikely.  High-dispersion spectroscopic observations
of these stars should clarify whether or not they are composite.  In the
meanwhile, it is probable that a re-evaluation of the physics on which 
current models are based may be required.

\section{Discussion}

It has been clear ever since the advent of high-precision photometry from
space, that there is a serious problem with current models of pulsating
stars in the upper main sequence.  Models of $\delta$~Sct stars predict that 
only frequencies higher than about 5\,d$^{-1}$ are unstable, yet all 
$\delta$~Sct stars, when observed with sufficient precision, show unexplained 
multiple low frequencies \citep{Balona2014a, Balona2015d}.  In addition, 
there is a growing realization that other types of pulsating stars exist
in the upper main sequence, besides the $\delta$~Sct, $\beta$~Cep and SPB 
variables.  Furthermore, there is evidence that the atmospheres of A and B
stars are more complex than previously thought.  Here we refer to
indications of superflares on A stars \citep{Balona2013c} and starspots on
both A and B stars \citep{Balona2016a}.

We need to remember that a large fraction of stars will show variability 
at the rotational frequency.  In the past, the presence of spots or
co-rotating features in upper main sequence stars has been thought unlikely
because radiative envelopes cannot support a magnetic field.  This view
needs to be reconsidered in the light of the fact that nearly half of all A
and B stars show what appears to be rotational modulation when observed with
sufficient photometric precision \citep{Balona2013c, Balona2016a}. 
Mis-classification of rotational modulation as pulsation certainly adds to
the confusion.

In this paper we describe two types on anomalous pulsating stars. The hot
$\gamma$~Dor variables have multiple low frequencies, but no high
frequencies, and lie between the red edge of the SPB and the blue edge of the
$\gamma$~Dor instability strip.  The Maia variables have multiple high
frequencies and lie between the red edge of the $\beta$~Cep and the blue
edge of the $\delta$~Sct instability strips.  Current models are unable to
account for pulsations in either of these two classes, adding to the
problems of pulsation in the upper main sequence.

We observed several hot $\gamma$~Dor candidates spectroscopically to
verify their effective temperatures.  In most cases the results confirm the
anomalous nature of these objects.  Two Maia candidates were also observed.
These results, together with the anomalous cluster stars observed by
\citet{Mowlavi2013} and \citet{Lata2014} and several well-observed brighter
stars, leave little doubt that these two kinds of anomalous pulsating stars 
do exist.

One possible explanation for the hot $\gamma$~Dor stars is that they
are rapidly-rotating SPB stars, as suggested by \citet{Salmon2014} and 
\citet{Balona2015c}. In a rapidly-rotating star, the equator is darker than 
the poles.  If the star is observed equator-on, it will appear cooler than if 
it is observed pole-on.  It is conceivable that such stars will appear to lie 
outside the red edge of the SPB instability strip and be seen as hot
$\gamma$~Dor variables.  In that case, all hot $\gamma$~Dor variables should 
have large projected rotational velocities, $v \sin i$.  The mean $v \sin i$ 
for B5--B9 main sequence stars (the typical spectral type range of SPB 
variables) is $\langle v \sin i\rangle = 144$\,km\,s$^{-1}$ as derived from 
the catalogue of \citet{Glebocki2005}.  The mean value of the 6 stars 
in Table\,\ref{gdor} with known projected rotational velocities is $\langle v \sin i\rangle = 
114$\,km\,s$^{-1}$, which does not lend any support to this idea.  In any 
case, it is difficult to see how rotation can move the apparent location of 
an SPB star to a location near the hot end of the $\gamma$~Dor instability 
strip, where most of these stars appear to lie (see Fig.\,\ref{hrdiag}).
In fact, \citet{Salmon2014} found only a slight decrease in the apparent
effective temperature of rotating stars as a result of gravity darkening.

A similar explanation could be proposed for the Maia variables, i.e. that
they are rapidly-rotating $\beta$~Cep stars observed nearly equator-on.
The mean projected rotational velocity for O9--B3 main sequence stars (the
typical spectral type range of $\beta$~Cep variables) is $\langle v \sin
i\rangle = 140$\,km\,s$^{-1}$ as derived from the catalogue of
\citet{Glebocki2005}.  On the other hand, the mean value for the 
7 stars in Table\,\ref{maia} is 92\,km\,s$^{-1}$, which does not support this 
idea.  Furthermore, it is difficult to understand how rotation can make a 
O9--B3 $\beta$~Cep star to appear as an late B or early A star, as many Maia 
variables appear to be.

One can also assume that these stars are composite, which was proposed by 
\citet{Balona2011b}.  In the case of a hot $\gamma$~Dor star the system
would consist of a non-pulsating early A or B star and a cool
$\gamma$~Doradus companion.  Because of the large luminosity difference,
it would be difficult to detect the $\gamma$~Dor companion in the composite 
spectrum.  This possibility certainly should be investigated using
high-dispersion, high signal-to-noise, spectroscopy.  We found no evidence of 
a cool companion in any of the stars that were observed.  A similar explanation
can be proposed for Maia variables.  In that case they need to be 
$\beta$~Cep stars with a cool, though fairly luminous non-pulsating companion 
or else a $\delta$~Sct star with a B-type companion.   This explanation does 
not seem plausible in the light of the numerous Maia candidates found in 
NGC\,3766 (as many as 20\,percent of the stars in the corresponding
magnitude range, \citealt{Mowlavi2013}) and in NGC\,1893 \citep{Lata2014}.

We conclude that the most probable explanation should be sought in a
revision of opacities \citep{Colgan2016, Moravveji2016}.  In the light of
the complexity of early-type atmospheres, a temperature inversion in the
upper atmosphere may exist.  In this case a modification of the outer boundary
condition in pulsating models should perhaps be considered.   Finally, it
might be worth exploring the possibility of sub-surface convection zones in
intermediate mass stars as was done by \citet{Cantiello2009} for massive
stars. 

\section*{Acknowledgments} 

This work has been done under the Indo-South Africa project DST/INT/SA/P-02.

This paper includes data collected by the {\it Kepler} mission. Funding for the
{\it Kepler} mission is provided by the NASA Science Mission directorate.
The authors wish to thank the {\it Kepler} team for their generosity in
allowing the data to be released and for their outstanding efforts which have
made these results possible.  

Much of the data presented in this paper were obtained from the
Mikulski Archive for Space Telescopes (MAST). STScI is operated by the
Association of Universities for Research in Astronomy, Inc., under NASA
contract NAS5-26555. Support for MAST for non-HST data is provided by the
NASA Office of Space Science via grant NNX09AF08G and by other grants and
contracts.
 
LAB wishes to thank the South African Astronomical Observatory and the
National Research Foundation for financial support.  SJ and ES acknowledges 
the grants INT/RFBR/P-118 and RFBR Grant No. 12-02-92693-IND\_a jointly 
funded by DST Govt of India and Russian Academy of Science, Russia.
Observations on the SAO RAS 6-meter telescope are carried out with financial
support from the Ministry of Education and Science of the Russian Federation
(agreement No. 14.619.21.0004, project ID RFMEFI61914X0004). ES is thankful
to the Russian Science Foundation (grant No. 14-50-00043) for financial
support of this study.

\bibliographystyle{mn2e}
\bibliography{hotgdor}

\label{lastpage}

\end{document}